# On the violation of the CHSH network model inequality. [1]

J.F. Geurdes
C.vd. Lijnstraat 164
2593 NN Den Haag, Netherlands
han.geurdes@gmail.com

Abstract
In this paper it is demonstrated that the quantum correlation (2-dim unitary parameter vectors) can be arbitrarily close approximated with a local hidden variables model. Moreover, the CHSH inequality can be violated with the model. This does not conclusively demonstrate that locality and causality can be restored to quantum mechanics. However, it does show that an experimental search for local hidden causality need not be fruitless beforehand.

1. Introduction

The early beginnings of quantum theory (QM) are marked by questions and debate of interpretation. In a later stage, an important step in the debate was made by Bell[1].

Based on Einstein's criticism of completeness[2], Bell formulated an expression for the relation between distant spin measurements such as described by Bohm[3]. In Bell's expression, hidden variables to restore locality and causality to the theory (LHV's) are introduced through a probability (mass) density function and through their influence upon the elementary measurement functions in the two separate wings (denoted by the A- and the B-wing) of the experiment.

Many experiments and theoretical developments arose from Bell's original paper. The most important experiment was performed by Aspect[4]. Aspect's results were interpreted as a confirmation of the completeness of quantum mechanics. From that point onwards, QM was considered a non-local theory. In a previous paper, the present author has discussed this proof[5] arguing that there was insufficient ground for such a far reaching conclusion.

First let us shortly describe a typical idealized Bell experiment. In such an experiment, from a single source, two particles with opposite spin are send into opposite directions. For instance, we could think of a positron and an electron arising from para-Positronium that are drawn apart by dipole radiation. Subsequently, in the respective wings of the experimental set-up, the spin of the individual particle is measured with a Stern-Gerlach magnet. The measurements are found to be correlated and depend on the, unitary, parameter vectors, $(a_1,a_2,a_3)$ and $(b_1,b_2,b_3)$ of the magnets. Quantum mechanics predicts the correlation to be

$$P(\vec{a},\vec{b}) = -\sum_{k=1}^{2} a_k b_k \tag{1}$$

The CHSH Bell inequality states that local hidden variables cannot recover P(a,b)=-(a.b), because all local hidden distributions run into inequalities similar to the one stated below.

$$D = P_{1,1} - P_{2,1} - P_{1,2} - P_{2,2} < 2\sqrt{2} \tag{2}$$

In the present paper, the contrast in Eq. 2 will be inspected more closely and it will also be shown that an LHV model is, like quantum mechanics, able to produce D≥2√2, i.e. to violate the inequality, D<2√2, in a numerical computer network model.

---

[1] The author would like to thank a friendly mathematician for his patience in discussing the matter with the author.

Looking at the, D, in Eq. (2), the question is justified whether the quantum mechanical correlation $P_{QM}(a,b)=-(a.b)$, is able to violate. For brevity we write, $a=(a_1,a_2,a_3)$ and $b=(b_1,b_2,b_3)$.

The first thing that perhaps may strike the reader is the fact that $P_{QM}(a,b)$ is symmetric in the arguments. Hence we have, $P_{QM}(a,b)=P_{QM}(b,a)$. Secondly, it follows that the 1 and 2 index in the D expression do not imply 1=a and 2=b, because then, it is easy to see that $P_{1,2}=P_{2,1}$ and because, in that case, $P_{1,1}=P_{2,2}=-1$, D=2 for every choice of unitary $a=(a_1,a_2,a_3)$ and $b=(b_1,b_2,b_3)$. Hence, in the restricted sense of the indices, i.e. 1=a and 2=b, the quantum correlation cannot violate either the inequality and there is no contrast between a computer network model for LHVs and QM. In other words, the D is a useless contrast in the sitation where A and B may randomly select from the set of unitary parameters, {a,b}, with a and b any choice of vector, $a=(a_1,a_2,a_3)$ and $b=(b_1,b_2,b_3)$, $|a|^2=|b|^2=1$.

Hence, the previous analysis implies that the indices in, D, need to be re-interpreted. We have,

$$D = P_{1_A,1_B} - P_{2_A,1_B} - P_{1_A,2_B} - P_{2_A,2_B} \qquad (3)$$

Hence, A may randomly select from unitary, {a,b} and B from the set {c,d} and, {a,b}≠{c,d}. It should be noted that the previous inequality can be violated with only a limited set of choices for {a,b}≠{c,d}. In the to be described computer experiment, we restrict the attention to a two-dimensional sub-space and take

$$a=(1,0), \qquad b=(0,1)$$
$$c=(1/\sqrt{2},-1/\sqrt{2}), \quad d=(-1/\sqrt{2},-1/\sqrt{2})$$

In the sections below, the "model" of local hidden variables will be outlined. The reader may also find this in ref 1 but here the formalism is rewritten to enable more easily the violation of the CHSH, D<2√2, in a computer network model such as described in Ref 2.

2. Preliminary remark on the form of the P(a,b) related to the $P_{i,j}$ from the D contrast

In this section the proposed model for a local hidden variables violation of the D contrast of type Eq. (2) derived from the EPR paradox is introduced. Firstly, let us note that Bell wrote the local hidden variables correlation function as

$$P_{LHV}(\vec{a},\vec{b}) = \int d\lambda \, \rho(\lambda) \, A(\vec{a},\lambda) \, B(\vec{b},\lambda)$$

with ρ(λ) the density of the hidden variables, λ, that are supposed to explain the correlation. Moreover, let us investigate the two-dimensional case with $A(a_1,a_2,\lambda)$ and $B(b_1,b_2,\lambda)$ where both element of {-1,1} are the spin-measurement functions of the A- and B- wing of the computer network model experiment (see Ref 2).

### 3. Measurement functions

Subsequently, sign weighted sums are introduced for both the A- as well as the B-wing.
For the ease of the presentation of the model, the probability densities of the different $\lambda$-type variables, are introduced later on in the paper. Most of the time the informed reader can already guess them from the definitions below.

A selects from $\{a,b\}$, while B selects from $\{c,d\}$. For completeness, $a=(a_1,a_2)$ and $b=(b_1,b_2)$ etc and unitary. In the first place let us define the form of the, to be used A and B functions that both project in $\{-1,1\}$. We have

$$A = sign\{\xi_A - \mu_A\}$$
$$B = -sign\{\xi_B - \mu_B\} \tag{4}$$

Note that $sign(x)=1$, for $x \geq 0$, and, $sign(x)=-1$, for $x<0$. The details of the probability density of, $\mu_A$, and, $\mu_B$, will follow in a subsequent section.

In the second place, let us continue the defintion of the structure of the hidden variables and present the two $\xi$ variables as

$$\xi_A(a) = \frac{a_1 S_1 + a_2 S_2}{\sqrt{2}} \sin(\alpha_A t_A) A_0$$
$$\xi_B(b) = \frac{b_1 S_1 + b_2 S_2}{\sqrt{2}} \sin(\alpha_B t_B) B_0 \tag{5}$$

Here, the $S_1$ and $S_2$, both in $\{-1,1\}$ are hidden variables that are shared by the A-wing and the B-wing. In the computer model of ref 2, the source computer will send $(S_1, S_2)$ to both A- and B-wing.

The probability densities for $t_A$ and $t_B$ will be given in a different section. Moreover, the value for the $\alpha$ parameters will be presented in the section where $-(a.b)$ will be derived from the model. Furthermore we have for $A_0$ and $B_0$ from Eq. (5)

$$\begin{Bmatrix} A_0 \\ B_0 \end{Bmatrix} = \begin{Bmatrix} 1, & \nu_{\{^A_B\}} \in [-N_{e\{^A_B\}}, 1) \\ sign\{\varphi_{N_{\{^A_B\}}, T_{\{^A_B\}}}(x_{\{^A_B\}}) - \nu_{\{^A_B\}}\}, & \nu_{\{^A_B\}} \in [1, N_{e\{^A_B\}}] \end{Bmatrix} \tag{6}$$

with,

$$N_{e\{^A_B\}} = \frac{N_{\{^A_B\}}}{1 - \frac{1}{1+(N_{\{^A_B\}}/\epsilon_{\{^A_B\}})^2}} \geq N_{\{^A_B\}} \tag{7}$$

$$1 \leq \varphi_{N_{\{^A_B\}}, T_{\{^A_B\}}}(x_{\{^A_B\}}) \leq N_{e\{^A_B\}}, \quad \forall x_{\{^A_B\}} \in [-T^2_{\{^A_B\}}, 1]$$

The definition of the density of, $\nu_A$ and $\nu_B$, will be given below. The definition of the indexed N values is related to this density.

In accordance with the restrictions given in (7), the $\varphi$ parameter function from Eq. (6) is defined.

We have

$$\varphi_{N_{\{B\}^A}, T_{\{B\}^A}}(x_{\{B\}^A}) = \begin{cases} (x_{\{B\}^A})^{-1}, & x_{\{B\}^A} \in [\tau_{\{B\}^A}, 1] \\ (1+T^2_{\{B\}^A})/2, & x_{\{B\}^A} \in [0, \tau_{\{B\}^A}) \\ 1, & x_{\{B\}^A} \in [-T_{\{B\}^A}, 0] \end{cases} \quad (8)$$

where, the τ-functions can be given as

$$\tau_{\{B\}^A} = \exp[-T^2_{\{B\}^A}(\sqrt{2}-1)] \quad (9)$$

Perhaps the reader is wondering why he should work his way through all the definitions. This question is justified, but it should also be remembered that only heuristics may guide an author in this situation because there are no physical reasons known for choices that were made. Having said that it is however easy to see the reason for the introduction of the, S, variables. For,

$$\sum_{S_1 \in \{-1,1\}} \sum_{S_2 \in \{-1,1\}} (a_1 S_1 + a_2 S_2)(b_1 S_1 + b_2 S_2) \propto a_1 b_1 + a_2 b_2 \quad (10)$$

and

$$\sum_{S_{\{2\}^1} \in \{-1,1\}} S_{\{2\}^1} = 0, \quad (11)$$

The other variables are necessary for purging this information out of sign function while remaining in the limits of classical probability (also to be found in Kolmogorov's axioms).

One other thing is also clear however, we have, $\xi_A$ and $\xi_B$ both projecting in the interval [-1,1].

In the network model (Ref 2) use is made of the raw product moment correlation, which depent on countings in certain combinations of settings, i.e.

$$\hat{R}(a,b) = \frac{N^=(a,b) - N^{\neq}(a,b)}{N^=(a,b) + N^{\neq}(a,b)} \qquad (12)$$

Here, $N^=(a,b)$ indicates the number of times, A=B occured, when the setting pair, a, and, b, was employed, etc. Note that, A, and, B, are the +1/-1 response that the measurement functions return. In the paper, there will be a relation established between the raw product moment correlation and the 'expectation' values in the model.

In order to start giving more clarity in the heuristically defined ranges of some variables, let us introduce the following equality between some parameters.

$$N_{\{^A_B} = \frac{1}{\epsilon_{\{^A_B}},$$

$$T_{\{^A_B} = \sqrt{\frac{\log(N_{\{^A_B})}{(\sqrt{2}-1)}} \qquad (13)$$

Sometimes it is handy to use either one of those in the presentation of the model. Basic to the expressions is the fact that one may take both the A-wing as well as the B-wing, ε model parameters, as a positive number as small as one pleases.

The relation in Eq. (13) warrants that the A-wing φ function from Eq. (8) remain in the interval $[1, N_{\epsilon\,A}]$ and likewise the B-wing φ function from Eq. (8) remain in the interval $[1, N_{\epsilon\,B}]$. This is necessary for a proper evaluation of both, $A_0$, and, $B_0$. Note in particular that, suppressing the indices for this moment, the φ have their maximum value, $\exp[T^2(\sqrt{2}-1)]$, which equals, N. Moreover,

$$(\forall x \in [0,\tau] \; \exists N_{\min} \; \forall N > N_{\min}) \; \varphi = \frac{1}{2}(1+T^2) = \frac{1}{2} + \frac{\log(N)}{2(\sqrt{2}-1)} \leq N_{\epsilon}$$

Hence, the sign expression in Eq. (6) is valid because (7) is implemented.

4. Probability model

Subsequently it is time to turn to the probability side of the advanced model. In the first place, let us define the probability density of the, x, variables in the model. We have

$$\rho_{x_{\{^A_B}}}(x_{\{^A_B}) = \begin{cases} (1+T^2_{\{^A_B})^{-1}, & x_{\{^A_B} \in [-T^2_{\{^A_B}, 1] \\ 0, & elsewhere \end{cases} \qquad (14)$$

Secondly, we define the t density as

$$\rho_{t_{^A_B}}(t_{^A_B}) = \begin{cases} 2t_{^A_B} N_{^A_B} e^2_{^A_B} / (t^2_{^A_B} N^2_{^A_B} + e^2_{^A_B})^2, & t_{^A_B} \in [0,1] \\ 0, & elsewhere \end{cases} \quad (15)$$

Thirdly, the A-wing and B-wing ν density is equal to

$$\rho_{v_{^A_B}}(v_{^A_B}) = \begin{cases} 1/2, & v_{^A_B} \in [-N_{e_{^A_B}}, N_{e_{^A_B}}] \\ 0, & elsewhere \end{cases} \quad (16)$$

Fourthly, the A-wing and B-wing μ density is equal to

$$\rho_{\mu_{^A_B}}(\mu_{^A_B}) = \begin{cases} 1/2, & \mu_{^A_B} \in [-1,1] \\ 0, & elswhere \end{cases} \quad (17)$$

As the stacked notation indicates, the previous variables reside separately on either the A-wing or the B-wing. The following two-dimensional spin type functions are shared by both the A- and the B-wing. We have,

$$\rho_{S_1,S_2}(S_1, S_2) = \frac{1}{4} \quad (18)$$

Subsequently, the A-wing and B-wing density is defined in a stacked A and B manner.

$$\rho_{^A_B} = \rho_{x_{^A_B}} \rho_{t_{^A_B}} \rho_{v_{^A_B}} \rho_{\mu_{^A_B}}$$

From the 'winged' definition, we can subsequently define the total density, $\rho_{tot}$.

$$\rho_{tot} = \rho_{S_1,S_2} \rho_A \rho_B \quad (20)$$

There is not much discussion necessary to acknowledge that, excluding for this moment the combined densities for, t, and for, ν, the introduced densities are perfectly 'classical' Kolmogorovian densities on the real domain. One can find them in any basic standard statistics textbook[6]. Special attention is given to the combined density of, t, and, ν.

Let us first introduce three lemmas that will be useful later in the presentation of the argument. Lemma-1 will be used in proving the unitarity of the hidden variables, 'λ', probability density, while the second Lemma will be used in the evaluation of the correlation. The third Lemma is necessary to establish the Kolmogorovian nature of the combined density of, t, and, ν.

*Lemma-1*    The following integral, $I_N$, over the t variable, is equal to $1/N_\varepsilon$.

$$I_N = \int_0^1 \frac{2tN\varepsilon^2}{(t^2N^2+\varepsilon^2)^2} \, dt$$

*Proof:*
Firstly let us substitute, $\omega = tN$. This gives the expression.

$$I_N = \frac{1}{N} \int_0^N \frac{2\omega\varepsilon^2}{(\omega^2+\varepsilon^2)^2} \, d\omega$$

From this equation it follows that

$$I_N = -\frac{1}{N} \int_0^N \frac{d}{d\omega} \frac{\varepsilon^2}{(\omega^2+\varepsilon^2)} \, d\omega$$

$$= \frac{1}{N}\left(1 - \frac{1}{1+(N/\varepsilon)^2}\right) = \frac{1}{N_\varepsilon}$$

which concludes this Lemma.

A second Lemma is also related to the, t, variable.

*Lemma-2*    The following integral, $J_N$, over the t variable and $\sin(\alpha t)$, 'goes to', 1.

$$J_N = \int_0^1 \frac{2tN\varepsilon^2}{(t^2N^2+\varepsilon^2)^2} \sin(\alpha t) \, dt$$

*Proof:*
Similar to the previous case, let us substitute, $\omega = tN$. This gives the expression.

$$J_N = \frac{1}{N} \int_0^N \frac{2\omega\varepsilon^2}{(\omega^2+\varepsilon^2)^2} \sin\left(\frac{\alpha}{N}\omega\right) d\omega$$

From this equation it follows that

$$J_N = -\frac{1}{N} \int_0^N \sin\left(\frac{\alpha}{N}\omega\right) \frac{d}{d\omega} \frac{\varepsilon^2}{(\omega^2+\varepsilon^2)} \, d\omega$$

Hence,

$$J_N = -\left(\frac{\sin\alpha}{N}\right)\frac{\epsilon^2}{N^2+\epsilon^2}$$

$$+\frac{\alpha\epsilon}{N^2}\int_0^N \cos\left(\frac{\alpha}{N}\omega\right)\frac{d}{d\omega}\arctan(\omega/\epsilon)\,d\omega$$

This equation can be further broken down into

$$J_N = -\left(\frac{\sin\alpha}{N}\right)\frac{\epsilon^2}{N^2+\epsilon^2} + \frac{\alpha\epsilon}{N^2}\cos(\alpha)\arctan(N/\epsilon)$$

$$-\frac{\alpha\epsilon}{N^2}\int_0^N \arctan(\omega/\epsilon)\frac{d}{d\omega}\cos\left(\frac{\alpha}{N}\omega\right)d\omega$$

Because, ε, may be chosen as a small positive real number and noting Eq. (10), arctan(N/ε) will approach, π/2. Similarly, the function, arctan(ω/ε), will be reasonably well be approximated with, π/2. From these considerations, together with Eq. (10) and selecting, α=2N³/π, we see that, $J_N=1+O(\epsilon^5)$, which concludes this second Lemma.

The third Lemma establishes the classical probabilistic nature of the model. We have,

*Lemma-3*     *The following integral, $K_N$, over, t and, ν, projects in the interval [0, 1].*

$$K_N = \int_\beta^\gamma dt\,\frac{2tN\epsilon^2}{(t^2N^2+\epsilon^2)^2}\int_{n_0}^{n_1}\frac{d\nu}{2},\quad with$$

$$[\beta,\gamma]\subseteq[0,1],\quad and\quad [n_0,n_1]\subseteq[-N_\epsilon,N_\epsilon],$$

$$\beta<\gamma,\quad n_0<n_1$$

*Proof:*
As can be easily verified there exists a, $N_0$, such that, for all, $N>N_0$

$$N \geq 1 + \frac{1}{N^4}$$

Hence, noting the relation between, N, and, ε, in Eq. (10), we have from the previous inequality

$$N^2 \geq \frac{N}{1-\dfrac{1}{1+(N/\epsilon)^2}} = N_\epsilon.$$

As can be seen from the integral expression for, $K_N$, we may evaluate

$$K_N = \int_\beta^\gamma dt \frac{2tN\varepsilon^2}{(t^2N^2+\varepsilon^2)^2} \int_{n_0}^{n_1} \frac{dv}{2}$$

$$= \frac{1}{2}(n_1-n_0)\frac{\varepsilon}{N^2}\frac{1}{\beta^2 N^2+\varepsilon^2}\left\{1-\frac{\beta^2 N^2+\varepsilon^2}{\gamma^2 N^2+\varepsilon^2}\right\}$$

Because of the previous inequality and the definition of the intervals, [β,γ], and, [$n_0$,$n_1$], it is easy to acknowledge that, for, 0<ε<1, the integral, $K_N$, projects in [0,1]. This concludes the third Lemma.

From this third Lemma and the first Lemma, it is already possible to acknowledge that the total density in Eq. (17) follows classical probability laws.

4. Evaluation of the model

In order to obtain the quantum correlation from the model, the expectation in relation to the total density is defined. We have,

$$E(F) = \sum_{S_1 \in \{-1,1\}} \sum_{S_2 \in \{-1,1\}} \int_{-T_A^2}^{1} dx_A \int_0^1 dt_A \int_{-N_{\varepsilon A}}^{N_{\varepsilon A}} dv_A \int_{-1}^{1} d\mu_A$$
$$\times \int_{-T_B^2}^{1} dx_B \int_0^1 dt_B \int_{-N_{\varepsilon B}}^{N_{\varepsilon B}} dv_B \int_{-1}^{1} d\mu_B \rho_{tot} F \tag{21}$$

In an intuitively understandable manner we will make use of the A- and B-wing expectation values and introduce this result into the overall expectation.

Let us define

$$E_{\{^A_B\}}(F_{\{^A_B\}}) = \int_{-T_{\{^A_B\}}^2}^{1} dx_{\{^A_B\}} \int_0^1 dt_{\{^A_B\}} \int_{-N_{\varepsilon\{^A_B\}}}^{N_{\varepsilon\{^A_B\}}} dv_{\{^A_B\}} \int_{-1}^{1} d\mu_{\{^A_B\}} (\rho_{\{^A_B\}} F_{\{^A_B\}}) \tag{22}$$

With the use of the previous definitions, and regarding Lemma-1, it is easy to see that, E(1)=1. Furthermore, let us inspect, $E_A(A)$. Hence, let us inspect

$$E_A(A) = \int_{-T_A^2}^{1} dx_A \int_0^1 dt_A \int_{-N_{\varepsilon A}}^{N_{\varepsilon A}} dv_A \int_{-1}^{1} d\mu_A \rho_A A \tag{23}$$

First we see, suppressing the A index here, that

$$\frac{1}{2} \int_{-1}^{1} d\mu \, sign\{ \frac{(a_1 S_1 + a_2 S_2)}{\sqrt{2}} \sin(\alpha t) A_0 - \mu \}$$
$$= \frac{(a_1 S_1 + a_2 S_2)}{\sqrt{2}} \sin(\alpha t) A_0 \tag{24}$$

Secondly, the, $\nu$, integral is, according to (6), equal to

$$\frac{1}{2} \int_{-N_{\epsilon_A}}^{N_{\epsilon_A}} d\nu_A A_0$$

$$= \frac{1}{2} \int_{-N_{\epsilon_A}}^{1} d\nu_A + \frac{1}{2} \int_{1}^{N_{\epsilon_A}} sign\{\varphi_{N_A, T_A}(x_A) - \nu_A\} d\nu_A \tag{25}$$

$$= \frac{1}{2}(1 + N_{\epsilon_A}) + \frac{1}{2} \int_{1}^{\varphi_{N_A, T_A}(x_A)} d\nu_A - \frac{1}{2} \int_{\varphi_{N_A, T_A}(x_A)}^{N_{\epsilon_A}} d\nu_a$$

$$= \frac{1}{2}(1 + N_{\epsilon_A}) + \varphi_{N_A, T_A}(x_A) - \frac{1}{2}(1 + N_{\epsilon_A}) = \varphi_{N_A, T_A}(x_A)$$

Thirdly, from Lemma-2 it follows that the integration

$$\int_{0}^{1} dt_A \rho_{t_A} \sin(\alpha t_A) = 1 + O(\epsilon_A^5) \approx 1 \tag{26}$$

for sufficiently small, $\epsilon_A$. So finally, the, $x_A$, integration must be evaluated.
We see, with the aid of previous equations, (7), (8), (9) that

$$\int_{-T_A^2}^{1} \frac{dx_a}{1 + T_A^2} \varphi_{N_A, T_A}(x_A)$$

$$= \int_{-T_A^2}^{0} \frac{dx_a}{1 + T_A^2} \varphi_{N_A, T_A}(x_A) + \int_{0}^{\tau_A} \frac{dx_a}{1 + T_A^2} \varphi_{N_A, T_A}(x_A) + \int_{\tau_A}^{1} \frac{dx_a}{1 + T_A^2} \varphi_{N_A, T_A}(x_A) \tag{27}$$

$$= \frac{T_A^2}{1 + T_A^2} + \frac{1}{2} \exp[-T_A^2(\sqrt{2} - 1)] + \frac{T_A^2(\sqrt{2} - 1)}{1 + T_A^2}$$

$$= \frac{\sqrt{2}}{1 + \frac{1}{T_A^2}} + O(\epsilon_A)$$

Combining the result of Eq. (22)-(24) and substituting in Eq. (20) gives,

$$E_A(A) \approx \sum_{k=1}^{2} a_k S_k \qquad (28)$$

Hence, applying the overall expectation we see that,

$$E(A) \approx \frac{1}{2} \sum_{S_1 \in \{-1,1\}} \frac{1}{2} \sum_{S_2 \in \{-1,1\}} (a_1 S_1 + a_2 S_2) = 0. \qquad (29)$$

Moreover, it is easy to see that, E(B), approximately is equal to zero as well. The overall expectation then can be written as

$$E(AB) = E_{S_1, S_2} E_A E_B (AB) = E_{S_1, S_2} (E_A(A) E_B(B))$$
$$\approx -E_{S_1, S_2} \sum_{k=1}^{2} a_k S_k \sum_{j=1}^{2} b_j S_j$$
$$= -\frac{1}{2} \sum_{S_1 \in \{-1,1\}} \frac{1}{2} \sum_{S_2 \in \{-1,1\}} (a_1 S_1 + a_2 S_2)(b_1 S_1 + b_2 S_2) \qquad (30)$$
$$= -\sum_{k=1}^{2} a_k b_k$$

In a more complete manner one may see the, ε, dependence,

$$E(AB) = -(\sum_{k=1}^{2} a_k b_k) \{\epsilon_A + \frac{1}{1 + \frac{\sqrt{2}}{|\log(\epsilon_A)|}}\}$$
$$\times \{\epsilon_B + \frac{1}{1 + \frac{\sqrt{2}}{|\log(\epsilon_B)|}}\} \{1 - \frac{\sin(\alpha_A) \epsilon_A^5}{1 + \epsilon_A^4}\} \{1 - \frac{\sin(\alpha_B) \epsilon_B^5}{1 + \epsilon_B^4}\} \qquad (31)$$

5. CHSH inequality

After having demonstrated the recovery of the quantum correlation from an LHV model it is necessary to show that the CHSH inequality can be violated in a computer network model. The computer network model restricts the possibilities of the proposed model to a small area of parameter settings and holds the strictest of possible conditions for a test of the truth of the proposed model.

In order to demonstrate the violation it is first necessary to outline the computer experiment more carefully. The selection of parameters for A and B is restricted to a small set as was already introduced in previous sections of this paper. Just below Eq. (3) the employed parameters are given and those parameters are such that D, in case of exact quantum mechanical correlation, is equal to $2\sqrt{2}$. For an approximation of the quantum correltion this perhaps is too small a violation but in principle, having the virtual possibility to approximate arbitrary close -(a.b), it must be possible.

In order to connect the model to the raw product moment correlations introduced in Eq. (12), we have to establish the following "expectation values" $E(A(u)B(w)|A=B)$ and $E(A(u)B(w)|A=-B)$. Let us start with, the A=B condition. More in particular with the A=B=+1 condition.

Note that one may get the impression that $E(A(u)B(w)|A=B=+1)$ is equal to unity. However, the conditional expression must be interpreted like is done in Aspect's paper (ref. 6), namely an integration over the set of hidden variables, given the parameter settings, that produces A=+1 and B=+1, etc.

In the model we see that A=+1 when, $\xi_A \geq \mu_A$ and B=+1 when, $\xi_B < \mu_B$. Hence, with this expectation, we have, for for instance the, a, vector (1,0) in the A-wing and the b vector, $(1/\sqrt{2}, 1/\sqrt{2})$ in the B-wing the following expectation

$$E(A(u)B(w)|A=B=+1)$$
$$= E_{S_1,S_2} E'_A E'_B [\frac{1}{4}(1+\xi_A(u))(1-\xi_B(w))] \qquad (32)$$

with E' meaning the expectation without the μ integration. The ξ containing factors in the previous equation can be explained as follows.

Under A=+1 looking at the $\mu_A$ integration only

$$\frac{1}{2} \int_{-1}^{\xi_A(u)} d\mu_A = \frac{1}{2}(1+\xi_A(u)) \qquad (33)$$

Under B=+1 lookin at the $\mu_B$ integration only

$$\frac{1}{2} \int_{\xi_B(w)}^{1} d\mu_B = \frac{1}{2}(1-\xi_B(w)) \qquad (34)$$

with u from {a,b} and w from {c,d} and a, b, c, d such as below Eq. (3).

Moreover, if we allow ourselves some extra evaluation of integrals, for A=-1 we see

$$\frac{1}{2} \int_{\xi_A(u)}^{1} d\mu_A = \frac{1}{2}(1 - \xi_A(u)) \qquad (35)$$

together with B=-1

$$\frac{1}{2} \int_{-1}^{\xi_B(w)} d\mu_B = \frac{1}{2}(1 + \xi_B(w)) \qquad (36)$$

Returning to the expectation in Eq. (32), and with the knowledge of the previous evaluations, we find

$$E(A(u)B(w) \mid A=B=+1) \approx \frac{1}{4}(1 - (u \cdot w)) \qquad (37)$$

Similarly for A=B=-1

$$E(A(u)B(w) \mid A=B=-1) \approx \frac{1}{4}(1 - (u \cdot w)) \qquad (38)$$

which then leads to, $E(A(u)B(w)|A=B) \approx (1-(u.w))/2$. From the Eqs. (33)-(36) it then follows that $E(A(u)B(w)|A=-B) \approx (1+(u.w))/2$. Hence, if we take the difference of the A=B expectation with the A≠B expectation, we see

$$E(A(u)B(w) \mid A=B) - E(A(u)B(w) \mid A=-B) \approx -(u \cdot w) \qquad (39)$$

Hence, the raw product moment correlation as defined in Eq. (12) can be interpreted in terms of the model as

$$E(A(u)B(w) \mid A=B) \approx \frac{N^=(u,w)}{N^=(u,w) + N^{\neq}(u,w)} \qquad (40)$$

and

$$E(A(u)B(w) \mid A=-B) \approx \frac{N^{\neq}(u,w)}{N^=(u,w) + N^{\neq}(u,w)} \qquad (41)$$

Such that, the difference between the A=B and the A=-B expectation corresponds to the raw product moment correlation from the computer experiment in ref 2.

$$\hat{R}(u,w) \approx E(A(u)B(w) \mid A=B) - E(A(u)B(w) \mid A=-B) \qquad (42)$$

The fact that this must be true follows because, in order to have a fair D contrast, it is necessary that when the raw product moment correlation is equal to the quantum correlation, with proper, u, and, w, vectors, the D inequality is violated. This puts the 'approximative recovery of the quantum correlation from an LHV model' on equal footing with 'the structure obtained from the model is able to violate the CHSH contrast in a computer setting'.

6. Conclusion

In the present paper, it is demonstrated that the quantum correlation can, with any desired precision necessary, be approximated with a local hidden variables model. This does not conclusively demonstrate that locality and causality can be restored to the quantum theory. However, it puts a question mark at all conclusiveness claiming demonstrations that locality and causality is eliminated with physical experiment.

The reader should note that distributions are not employed in the present model. Moreover, it is sufficient to approximate the qm correlation and, in this way, to enable a numerical simulation of the qm correlation in a computer. The key to this is the approximation of the raw product moment correlation from the computer experiment and the conditioned expectations from the model.

References

[1.] Bell, J.S. (1964) On the Einstein Podolsky Rosen paradox. *Physics*, **1**, 195-200.

[2.] Einstein, A., Rosen, N. and Podolsky, B. (1935) Can quantum-mechanical description of physical reality be considered complete? *Phys. Rev.*, **47**, 777-780.

[3.] Bohm, D. The Paradox of Einstein, Rosen and Podolsky. In: Quantum Theory and Measurement, page 354-368. Edited by J.A. Wheeler and W.H. Zurek, Princeton Univ. Press, 1983.

[4.] Aspect, A., Dalibard, J. and Roger, G. Experimental test of Bell's inequalities using time-varying analyzers. *Phys. Rev. Lett.*, **49**, 1804-1806.

[5.] Geurdes, J.F. (1998 a) Quantum Remote Sensing, *Physics Essays*, 11, 367-372.
Geurdes, J.F. (1998 b) Wigner's variant of Bell's inequality, *Austr. J. Phys.*, 51(5), 835-842.
Geurdes, J.F. (2001) Bell inequalities and pseudo-functional densities, *Int. J. Theor. Phys., Group. Theor. & Nonl. Opt.*, 7(3), 51.
Geurdes, J.F. (2006) A Counter-Example to Bell's theorem with a softened singularity, *Galilean Electrodynamics,* 17(1), 16.
Geurdes, J.F. (2007) Bell's theorem refuted with a Kolmogorovian counter-example, *Int. J. Theor. Phys., Group. Theor. & Nonl. Opt.*, 12(3), 215.

[6.] Mood, A. M., Graybill, F. A. & Boes, D. C., *Introduction to the theory of statistics*, Mac Graw-Hill, 1974.